# An Assessment Of UK Economy Through Payment Transactions Data

## Principles of Data Science

City University , London

arunav.das@city.ac.uk

Arunav Das

Abstract *UK GDP data is published with a lag time of more than a month and it is often adjusted for prior periods. This paper contemplates breaking away from the historic GDP measure to a more dynamic method using Bank Account, Cheque and Credit Card payment transactions as possible predictors for faster and real time measure of GDP value. Historic timeseries data available from various public domain for various payment types, values, volume and nominal UK GDP was used for this analysis. Low Value Payments was selected for simple Ordinary Least Square Simple Linear Regression with mixed results around explanatory power of the model and reliability measured through residuals distribution and variance. Future research could potentially expand this work using datasets split by period of economic shocks to further test the OLS method or explore one of General Least Square method or an autoregression on GDP timeseries itself*.

## I. INTRODUCTION

GDP is a measure of health of an economy and an important indicator for macroeconomic policy decisions related to setting the borrowing rates, taxation and government spending plans.

UK GDP is measured [1,2,10] as a combination of total value of good and services produced (Output Measure) by all sectors, total value of good and services bought (Expenditure Measure) by government and households and income related to profits and wages (Income Measure). GDP data is released on monthly and quarterly basis but with a lag of more than a month. They are often corrected and final versions are only available three months after quarter end.

As each component of GDP measure has an underlying economic activity that is settled through various domestic and international transactions made by companies, consumers, government, investors using bank account, cheque, cash, debit or credit cards payments, these transactions are assumed to be good predictors for GDP. Motivation of this study therefore is to explore correlation between payments and GDP to model an alternative GDP, forecasting model thereby potentially replacing the traditional way of measuring GDP.

Literature review [3] shows past/present focus on forecasting GDP has been on parameters like Inflation and Treasury Bills as well as defining new wellbeing measures of GDP [4,5]. Successful modelling of GDP with Payment Transaction could particularly be useful for wellbeing measures as the metadata content of underlying transactions could be analysed to understand type of transactions trending up or down in an economy

Machine Learning techniques provides ability to query historic datasets, mine trends, patterns, correlations for assessing explanatory power of features variables for future predictions. This research is an attempt to find and use an appropriate Payment transaction feature for Regression Modelling techniques to measure and predict UK GDP

## II. ANALYTICAL QUESTIONS & DATA

*A. Research questions*

- Do timeseries data for Bank Account and Card Payment Transactions exhibit similar characteristic like GDP timeseries data

- Is there a correlation between Payment Transaction timeseries data and GDP timeseries data

- Could a Simple or Multiple Linear Regression model be built using timeseries Payments data to predict GDP

- How would Linear Regression Model built on timeseries dataset compare to alternate model that specifically deal with timeseries datasets

*B. Datasets*

Timeseries datasets used for

1. Monthly High and Low Value, Cheque [13] and Credit Card payments [14] as Predictor variables

2. Quarterly UK GDP data as Target variable [15]

Boxplots of predictor variables (*Figs1,2*) visually demonstrates negatively skewed distribution for both volume and values with heavy tails and outliers for High Value and Credit Card Payments and thin tail for Cheque and Low Value Payments.

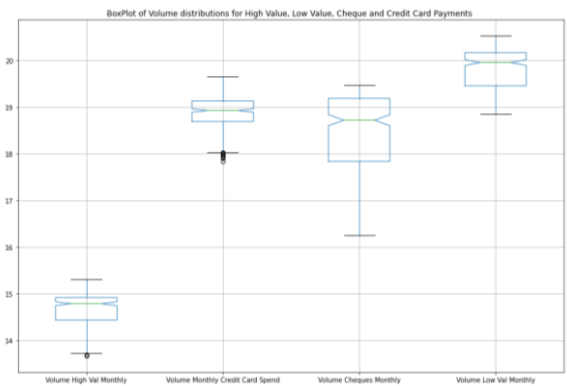

Fig1

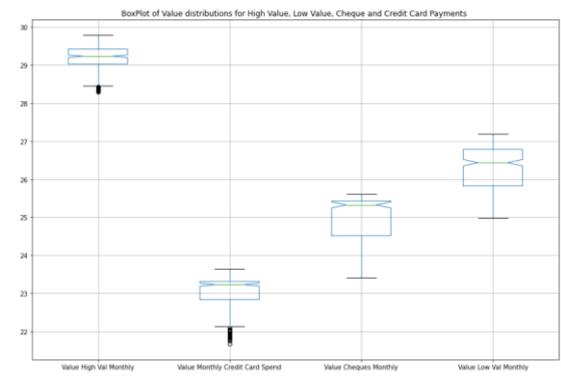

Fig2

However, descriptive statistics shows minor contradictions as Low Value Payments appear to have normal distribution and High Value and Credit Card Payment exhibit lower kurtosis i.e. thin tails instead of outliers. Other observations from descriptive statistics are aligned with visual observations

Decomposition[6] of predictor variables (Low Value Payments) shows (*Fig3*) timeseries data trending upward, quarterly seasonality and uniform distribution of residuals

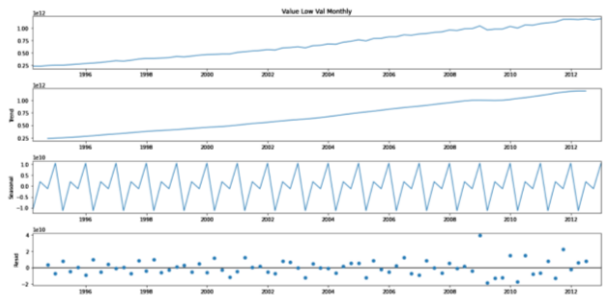

Fig3

Quarterly Boxplots *(Fig4)* of Nominal GDP data visually exhibits fairly symmetrical distribution

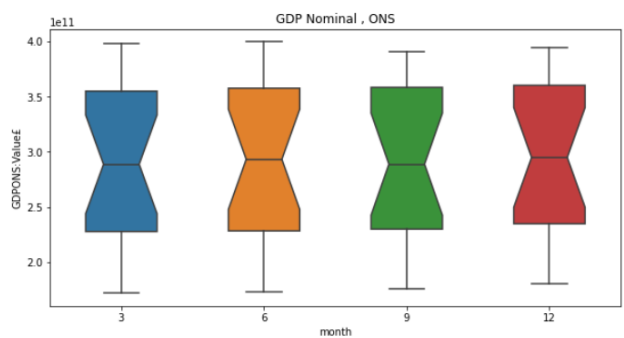

Fig4

Descriptive statistics of GDP show a symmetrical distribution with thin tails for Nominal and Inflation Adjusted GDP for ONS source. The US FED inflation adjusted GDP appear to be slight negative skew

Decomposition (*Fig5*) of GDP data show timeseries trending upward with quarterly seasonality. Distribution of residuals are fairly and densely even until 2008 with distributions become sparser and less regular afterwards

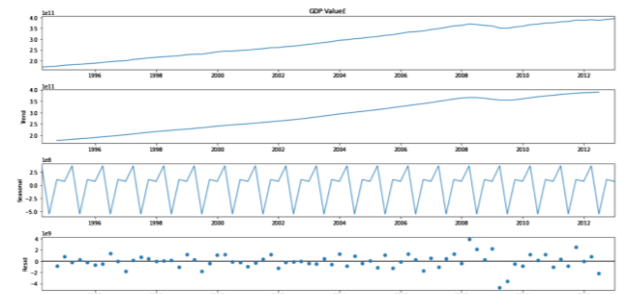

Fig5

An assumption has been made that statistical similarities between some of predictor variables with target variable implies source datasets could be used for regression modelling.

### III.  ANALYSIS

*A.  Data preparation*

Source data in various formats - multiple tabs excel for Payments and csv for GDP, was uploaded to create datasets with consistent format. Column headers were manually assigned as source files have a number of header and footer rows with narrative and also as column headers appear across multiple rows. Year value was imputed for missing rows. Year, Month were converted from Float , Object type respectively to Datetime. Transaction volume and values were assigned absolute values based on numerical conversions

GDP source file was corrected for formatting issues related to date fields with extra space that caused issues while uploading source dataset.

GDP dataset have quarterly timeseries whereas the payment datasets have a monthly timeseries. Datasets were accordingly uploaded.

Datatypes checked to ensure amounts are in float and time periods in datetime format consistently.

Initial data preparation was subsequently revisited to correct source file formatting issues identified through Scatter plot during the Exploratory Data Analysis steps. Payments (BACS) source file was corrected for a differently formatted amount from 2010 to 2020 causing overstatement by 10^3. Cheques file had data available in different columns between period June 2009 till Sep 2019 vs pre-June 2009 period. Following corrections, Timeseries (*Fig6*) plot confirms data quality and reliability [7]

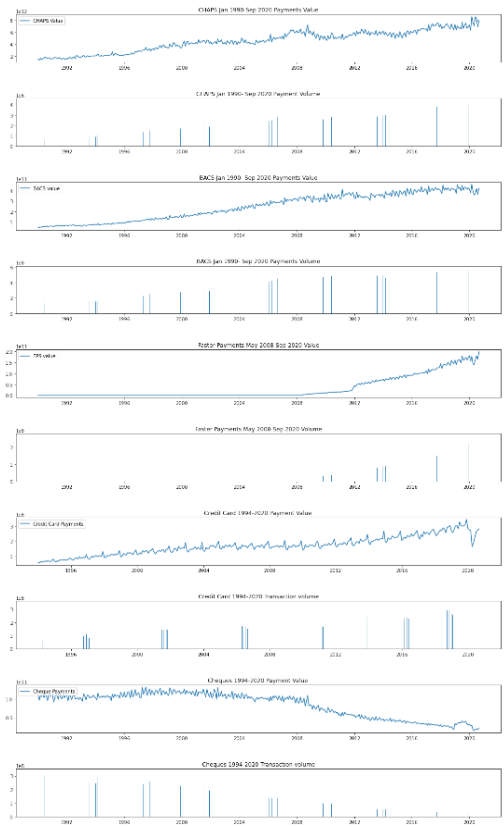

Fig6

### B. Data Derivation

Multiple datasets representing different types of bank account and credit card payment transactions were initially merged using timeperiod parameter, creating four unique Payments datasets – High Value, Low Value, Cheques and Credit Cards. New features were derived at this stage for merger of datasets, reducing number of fields for analysis from 18 transaction volume related features to 5 and 13 transaction value related features to 5.

Visual analysis was performed to ensure no distortion was caused by datasets merger and feature aggregation steps. This step resulted in formatting spotting anomalies (*Fig7a*) in source data that was corrected and reprocessed for validation (*Fig7b*)

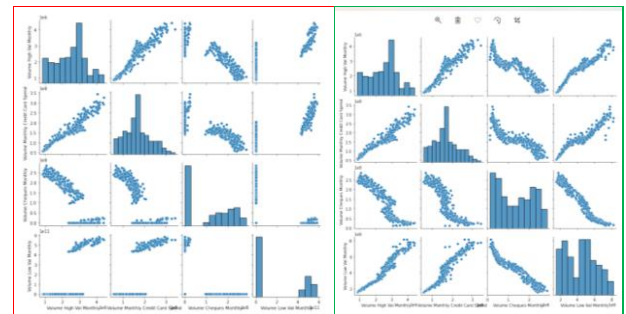

Fig7a       Fig7b

A new consolidated payments dataset was created from four datasets to carry out Exploratory Data Analysis. As the source datasets had different timeperiod, unification of datasets caused payments data prior to 1994 to be dropped.

As the GDP dataset has a quarterly timeseries, the payments dataset was converted into a quarterly timeseries as well.

As GDP value is a monetary measure of economic activities, it is assumed that neither the volume of transactions nor standalone value would be a useful predictor. Hence, new features were created for each payment type to represent average transaction values for High Value, Low Value and Credit Card payments for a combined comparative of transactionality and magnitude of payments data with GDP

### C. Construction of models

Cheque Payments [7] were not considered for modelling as it's use has been declining in UK and replaced by electronic payments. Hence, it may not be a useful predictor in the long run. Scatterplot (*Fig8*) visually shows Average transaction value for Low Value and Credit Card Payments have correlation with GDP whereas a distinct correlation is not visible for High Value Payments

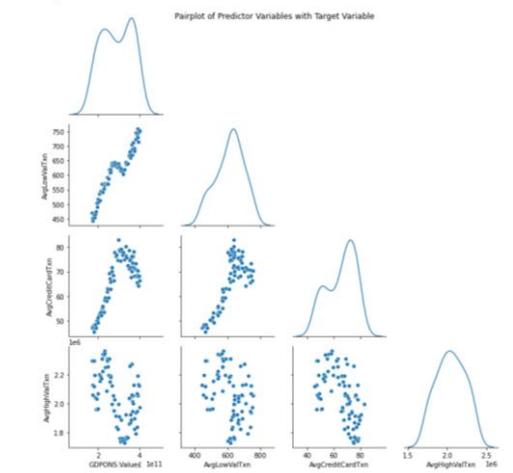

Fig8

Low Value Payments (0.95) was chosen as the Predictor variable over Credit Card (0.79) and High Value (-0.47) using the results from Correlation Matrix (*Fig9*)

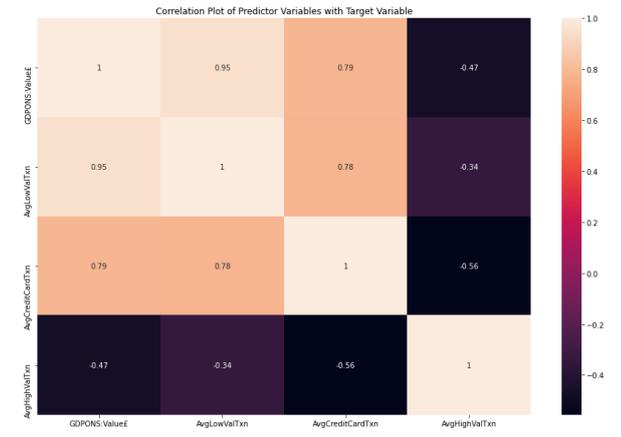
Fig9

High Variance Inflation Factor between the feature variables resulted in the choice of a simple Linear Regression model as opposed to Multiple Linear Regression (MLR).

Seaborn Regression plot (*Fig10*) was used to check the relevance of the Simple Linear Regression Model for the chosen predictor variable

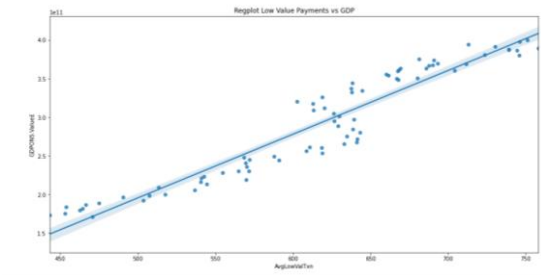
Fig10

Statsmodel OLS Simple Linear Regression model was built using 66% of the dataset for Training and 33% for Testing.

D. Model Evaluation

Per OLS Model summary (*Fig11*), 91.7% ($R^2$) of GDP variation is explained by Low Value Payments. Regression coefficients appear to be statistically significant though a negative coefficient should be checked further. F-statistics show that predictor-target relationship is meaningful

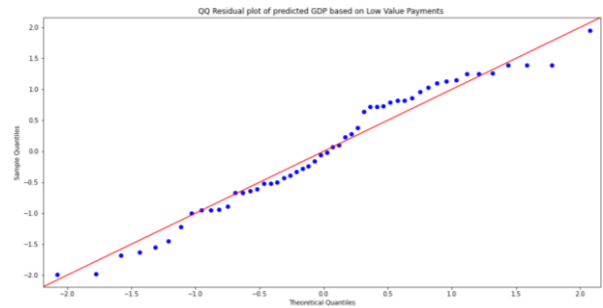
Fig11

Harvey-Collier multiplier test shows no violation of linearity. Predicted vs actual values show positive variance below mid-range GDP values and negative variance above. Influence plot shows presence of high leverage, low residual and low leverage, high residual points indicating outliers with influence on the model.

QQ plot (*Fig12*) of residuals [8] shows fairly symmetrical distribution, confirmed by Jarque-Bera (>.05). Normality therefore could be assumed. However, presence of tails in QQ plot, bimodal distribution of histogram plot (*Fig13*) of residuals and lower Omnibus probability needs to be investigated further to validate normality.

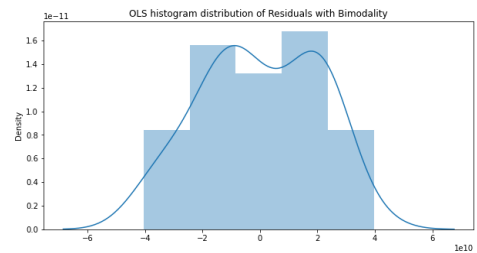
Fig12

Fig13

Homoscedasticity cannot be conclusively validated as residual vs fitted plot (*Fig14*) neither exhibit funnel, cone shape nor uniform distribution. The plot has "Z" shape implying overestimation at lower at higher predictor values and underestimation at midrange values

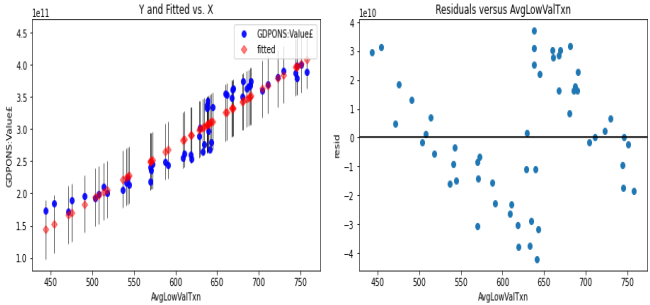

Fig14

High condition number for model potentially caused by timeseries autocorrelation with presence of discernible lags (Fig5) demonstrating lack of independence.

Model therefore demonstrates Linearity and reasonable Normality but no clear evidence of Homoscedasticity is observed. These validations require further analysis for concluding on model robustness [16]

*E.* Validation of results

Model was found to return consistent results from multiple runs with different training/test sample sizes. Model performance was found to be better for chosen deature than other two predictor variables. Multiple Linear Regression was used and not found to improve the results nor residuals.

General Least Squares Modeling was used as comparative to mitigate timeseries autocorrelation impact on OLS Linear Regression. GLS returned higher predictability and more robust coeffients. Distribution of residuals were found to have less autocorrelation but residual distribution was less nornal and more bimodal (*Figs15,16*). GLS shows a much higher explanatory power for other predictor variables compared to OLS model

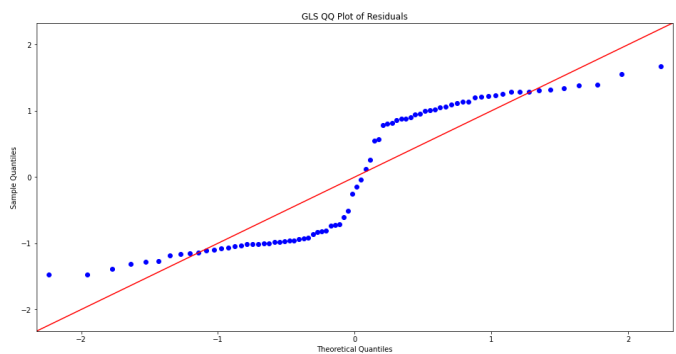

Fig15

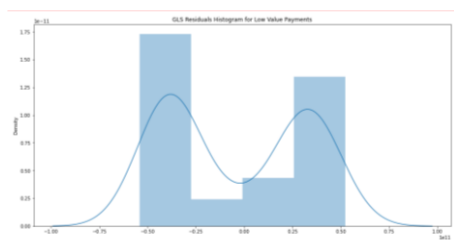

Fig16

Sklearn Linear Regression with Cross-Fold validation was used for checking overfitting and robustness under random sample size. While simple regression resulted in similar explanatory power (93%), Cross-Fold Validation (splits =2&3) shows significant variation of explanatory power between 81.8%-90%. Larger split were not used due to the small sample size (=78). Results indicate at higher explanatory power, OLS is potentially overfitted. Small sample size [17] could mean these results are not completely reliable.

Polynomial fit model does not improve fit nor variance (*Fig17*) from OLS Simple Regression though it improves normal distribution of residuals

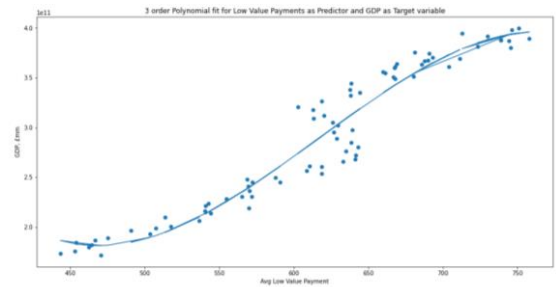

Fig17

## IV. FINDINGS , REFLECTIONS, FURTHER WORK (600 WORDS)

Analysis of Payment Types confirms non-stationary timeseries attributes, like GDP timeseries, with discernible trends, seasonality and residues. High correlation was found between various Payment Types and GDP though negative correlation with High Value payments is surprising and needs to be investigated further

Amongst the Payment Types considered, Low Value Payments appears to be best predictor for GDP under OLS Simple Linear Regression Model. Whilst explanatory power of this model is very high with robust coefficients, it is difficult to conclude OLS modelling approach is ideal because of mixed results obtained for residuals distribution and variance. Though residuals QQ plot shows near normal distribution with low skewness, it also exhibits high variance, bimodality and autocorrelation raising doubts about reliability of OLS beyond available datasets. Furthermore, high explanatory power shows an observable variance between 81% to 90% under Cross-Fold Validation method. This could point to overfitting issues with OLS model but the sample size (78) is very small to draw meaningful conclusions from Cross-Fold Validation.

As autocorrelation is problematic for OLS Regression [10,11], General-Least-Square Regression model was used

for countering timeseries autocorrelation issues. GLS model shows high explanatory power with better coefficient robustness and lower residual autocorrelation compared to OLS but it also fails to conclusively address bimodal and variance issues.

Influence Plots (*Fig18*) show very high sensitivity of OLS to few low leverage predictor values and outliers. High Condition Numbers in OLS summary reinforces model sensitivity. This sensitivity alongwith residual autocorrelation perhaps limits the OLS to short-term GDP forecasts.

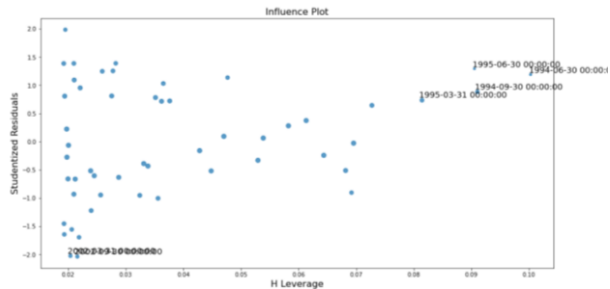
Fig18

One plausible cause for residuals bimodality and outlier sensitivity could be impacts from 2002 Dotcom crash and 2007-8 Financial Crisis. An extension of this research could be based on modelling datasets from different periods pre/post economic shocks to address bimodality and outlier sensitivity issues. Once resolved, model could be applied to predict post Covid19 economic forecasts due to similar economic shocks and impact on underlying payment volumes and values

Household consumption [6] accounts for 62% of Expenditure GDP. Its not surprising to see Low Value Payments with high correlation to GDP. However, the lower predictability of Credit Card Transactions in the OLS model (though higher for GLS model) should be investigated further. Also, future work should consider Debit Card and Cash Transactions as additional features.

High explanatory power of High Value, Low Value, Credit Card payments under the GLS method combined with very high correlation coefficients shows that GLS model should be explored further especially if splitting dataset as per pre/post economic shocks solves bimodal issue of residuals

As different types of payments are not substitutive in nature, driven by different economic activities and some payments are made using available credit while others using available balance in bank accounts, a further investigation of correlation and Variance Inflation Factor tests on split datasets could help to understand alternate feature engineering methods for MultiLinear Regression method.

Nominal GDP was used as the Target variable and perhaps the same modelling approach should be repeated to assess whether the predictors have better explanatory power for inflation adjusted GDP under OLS, GLS and MLR.

In summary, all three models – OLS, GLS, MLR should be investigated further with datasets split trained/tested by different economic periods of stability and shocks to explore if the bimodality and residual variance issues are addressed

Quarterly variance is key GDP measure and was not considered for this research. Other key variables excluded from this analysis are - impact of overseas transactions, transactions analysis by industry/sector, specific impacts on payment types from economic shocks and their correlation to GDP

Finally, whilst this research is based on explanatory power of payment transactions for GDP, GDP timeseries itself could be used as regressor to predict future values using an Autoregression model better suited for timeseries.